\documentclass[]{spie}

\title{Experimentally-determined performance limits for joint imaging and wavefront sensing with a photonic lantern}

\author[a]{Aditya R. Sengupta}
\author[c]{Vincent Chambouleyron}
\author[a]{Rebecca Jensen-Clem}
\author[a]{Emiel Por}
\author[b]{Benjamin L. Gerard}
\author[a]{Jordan Diaz}
\author[d]{Zoe Weber-Porter}
\author[e]{Yoo Jung Kim}
\author[a]{Steph Sallum}
\author[a]{Matthew DeMartino}
\author[a]{Daren Dillon}
\author[a]{Kevin Bundy}
\author[a]{Anna K. Gagnebin}
\author[a]{Philip Hinz}
\author[f]{Caleb Dobias}
\author[f]{Tara Crowe}
\author[f]{Stephen S. Eikenberry}
\author[f]{Rodrigo Amezcua-Correa}
\author[f]{Stephanos Yerolatsitis}

\affil[a]{Department of Astronomy \& Astrophysics, University of California, Santa Cruz, CA 95064, USA}
\affil[b]{Lawrence Livermore National Laboratory, Livermore, CA 94550, USA}
\affil[c]{Laboratoire d'Astrophysique de Marseille}
\affil[d]{School of Engineering, University of California, Santa Cruz, CA 95064, USA}
\affil[e]{Physics and Astronomy Department, University of California, Los Angeles, 475 Portola Plaza, Los Angeles, 90095, USA}
\affil[f]{CREOL, The College of Optics \& Photonics, University of Central Florida, Orlando, FL
32816, USA}

\authorinfo{Further author information: (Send correspondence to A.S.)\\A.S.: E-mail: adityars@ucsc.edu}

\usepackage{preamble}

\begin{document} 
\maketitle
\begin{abstract}
The photonic lantern (PL) is a focal-plane wavefront sensor (WFS) that has been used for second-stage control of extreme adaptive optics (AO) systems. While the number of sensed modes and the dynamic range with respect to each mode have been relatively well characterized, little attention has been paid to the PL's sensitivity, i.e. how measurement noise impacts the accuracy of PL wavefront reconstruction. We compute the PL's sensitivity to photon noise as a function of spatial frequency, and compare it to existing WFSs, using simulations as well as experiments on the muirSEAL testbed. We further assess these metrics in the case where only a subset of PL ports are available for wavefront sensing. In this configuration, the remaining ports are used to spatially and spectrally reconstruct the observed science scene using algorithms such as SPADE. Using more ports for wavefront sensing enables greater aberration sensitivity but leaves less spatial information for image reconstruction. This allows us to trade off between fewer samples with smaller aberrations and more samples with larger aberrations. This work sets the stage for AO system designs that incorporate the PL as a joint WFS and imager.
\end{abstract}

\keywords{Astrophotonics, photonic lantern, wavefront sensing, wavefront control, adaptive optics}

\section{Introduction}

Photonic lanterns (PLs) are fiber-based optical devices that taper from a multi-mode input end to several single-mode output ports, splitting input light apart based on its spatial pattern. This property allows them to be used as low-order focal-plane wavefront sensors (WFSs) \cite{Norris20,Lin22} while additionally directing light downstream to science instruments, correcting the non-common-path aberrations and adaptive optics residuals that currently limit high-contrast imaging. 

In the past few years, PLs have been employed for a wide range of purposes in astrophotonics. Recently, PLs as second-stage WFSs that correct non-common-path aberrations have been shown to improve image quality on sky. Lin \textit{et al.} 2025\cite{Lin25} demonstrated closed-loop control on a PL with dispersed outputs on the Subaru 8m telescope, and in Sengupta \textit{et al.} 2026 \cite{Sengupta26onsky}, we demonstrated closed-loop control on a PL on the Shane 3m telescope. In the latter work we filtered the PL output instead of dispersing it, making more of the spectrum available for science at the cost of correcting fewer aberration modes. Laboratory demonstrations have further shown that PLs are able to sense the low-wind effect\cite{Wei24} and primary-mirror segment offsets\cite{Cuevas25}. As science instruments, PLs with dispersed outputs have been used on sky for high-resolution spectroscopy\cite{Vievard24} and spectro-astrometry\cite{Kim25}; this effort recently led to a positional measurement of a disk with a photocenter precision of $\sim$50 $\mu$as. PLs have been used for nulling interferometry, both by using the lantern itself\cite{Xin24} and by feeding its outputs to a downstream interferometer\cite{Diaz26}. PLs have also been demonstrated in the lab as reconstructive spectrometers\cite{DeMartino26,Gagnebin26}, suggesting their spatial splitting property may also be applicable in spectral space.

An argument based in fundamental physics for why PLs should be expected to excel at many of these tasks is found in the theory of mode-based imaging. Tsang \textit{et al.} 2016\cite{tsangQuantumTheorySuperresolution2016} first demonstrated that linear optics could outperform the classical diffraction limit via the method of spatial-mode demultiplexing (SPADE), and Kim \textit{et al.} 2025\cite{Kim25} discusses photonic lanterns as direct implementations of SPADE. They show that ideal mode-selective lanterns have nonzero Fisher information with respect to the double-source resolving problem (e.g. imaging a binary), but zero Fisher information with respect to the single-source resolving problem (e.g. tip/tilt WFS). On the other hand, ideal standard lanterns achieve nonzero Fisher information for the single-source resolving problem, at the cost of being less sensitive to double-source and other higher-order patterns. SPADE is known to achieve the tighter resolution limit set by quantum information theory\cite{Huang21,Grace22}. This suggests that if ideal lanterns are perfect implementations of SPADE, rather than just noting lanterns are capable of sensing these scenes, we can investigate their optimality. We note that mode-based imaging outperforming classical imaging (in the sense of enabling imaging past the diffraction limit) does not imply a similar relationship would hold for WFS, and extending methods like SPADE to set constraints on WFS performance is a promising avenue for future work; however, PLs being direct implementations of SPADE is sufficient to motivate our investigation of the PL relative to classical fundamental WFS limits.

This empirical and theoretical promise invites us to consider relevant metrics for the PL, as an intermediate step towards making performance predictions for future photonic instruments. In this work we will focus primarily on WFS, but will consider its impact on imaging by assuming the science objective is to maximize the Strehl ratio for an image of a single source, and by assuming some ports are reserved for science imaging and cannot be used for WFS in real time. We chose this approach because we have only a standard (not mode-selective or hybrid) lantern available for laboratory experiments.

In this work, we conduct the first simulations and experimental measurements of photon noise sensitivity for the photonic lantern wavefront sensor. We investigate these in both a `pure-WFS' mode and a `dual imager/WFS' mode, in which we assume some subset of PL ports is unavailable for wavefront sensing. In Section 2, we introduce the relevant theory regarding PLs and sensitivity to photon noise. In Section 3, we present simulations of photon noise sensitivity for several PL designs at a range of focal ratios of the PSF at the PL entrance. In Section 4, we present experimental measurements of photon noise sensitivity from the muirSEAL testbed in the UCSC Laboratory for Adaptive Optics. Section 5 is a conclusion.

\section{Theory}

\subsection{Overview of photonic lantern wavefront sensing}

Photonic lanterns act as low-order wavefront sensors where the number of modes they can sense is limited by their number of output ports. We can model PLs by computing their basis of principal modes, i.e. the electric fields at the multi-mode end that excite each individual single-mode port in turn, and projecting the electric field of an input PSF onto this basis. Assuming a fixed PL design and a diffraction-limited PSF as the nominal input (which may not always be optimal; e.g. Lin \textit{et al.} 2023\cite{Lin23} consider the effect of beam-shaping optics on WFS capabilities), the remaining degree of freedom is the size of the input PSF, as set by the focal ratio. 

\begin{figure}[h!]
    \centering
    \includegraphics[width=\textwidth]{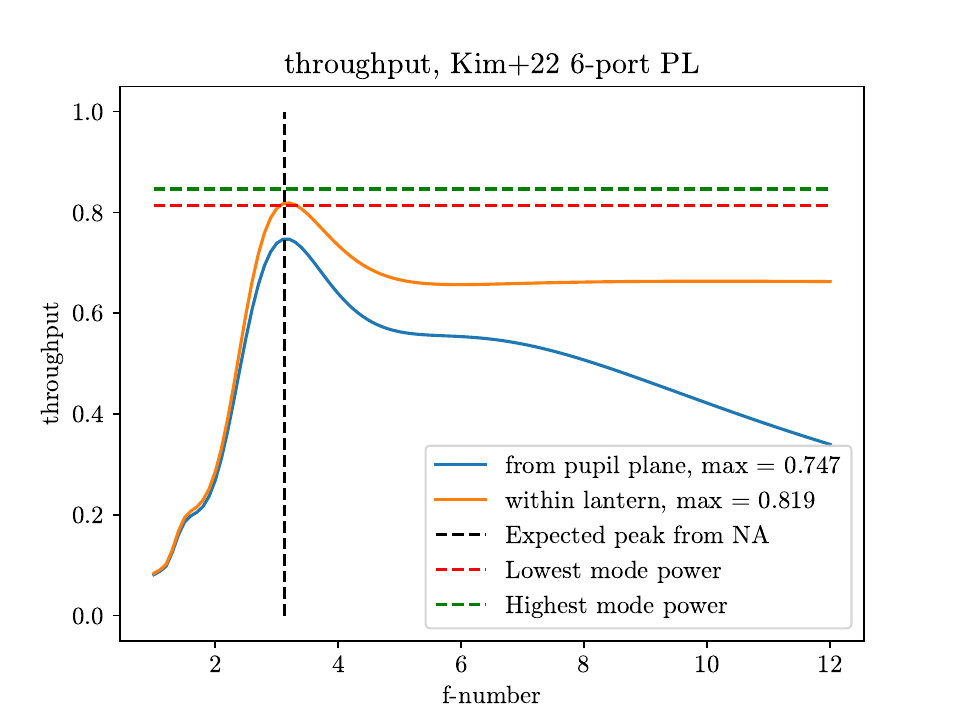}
    \caption{Simulated throughput as a function of f-number for a 6-port standard photonic lantern. The blue line is normalized to the total power in the pupil plane before being brought to a focus and coupling into the PL, and the orange line is normalized to the total power that couples into the multi-mode end of the PL. The peak of both curves occurs at the predicted point based on the numerical aperture of the multi-mode end. The red and green lines show the minimum and maximum power as calculated by \textit{lightbeam} carried by the principal modes, i.e. they characterize the most and least insertion loss per mode, and we see that the peak power for the orange curve falls in between these.}
    \label{fig:optimal_injection}
\end{figure}

PLs have an focal ratio that maximizes their throughput, which is set by the numerical aperture at the multi-mode end to be $f/\# = 1 / (2 \times \text{NA})$, assuming an adiabatic taper. The throughput as a function of focal ratio more generally is largely set by the injection efficiency, i.e. what fraction of light never couples into the lantern. For focal ratios less than the peak, insufficient light is coupled into the outer ports and more light is lost to the cladding; for focal ratios greater than the peak, light falls outside the multi-mode end and does not couple into the lantern. The resulting throughput curve is shown in Figure~\ref{fig:optimal_injection}, for the 6-port PL used in Kim \textit{et al.} 2022\cite{Kim22} (henceforth K+22).

Maximizing throughput seems to be a clear design goal, but not every metric of PL performance is optimized at the peak throughput. In particular, we will later see that underfilling the PL input makes it easier to sense higher-order aberrations.

\subsection{Motivation for photon noise sensitivity}

The primary metric that has been used until now in assessing PL WFS performance is dynamic range, i.e. the largest aberration amplitude such that a control loop can be successfully closed and the input wavefront can return to a flat state. Simulations of PL wavefront sensing on the low-order aberrations yield linear ranges of around 1 rad \cite{Lin22,Lin23}. Laboratory experiments found similar results of around 1-2 rad \cite{Sengupta24,Sengupta25}. These can be further extended for particular modes using nonlinear reconstruction algorithms\cite{Lin24fewmode,Sengupta24,Wei24}. Since virtually all AO systems are capable of delivering post-first-stage wavefronts within the linear range, even before factoring in extensions from nonlinear reconstruction, dynamic range is unlikely to be the limiting factor for the performance of a PL as a joint imager and second-stage WFS.

Therefore, we instead focus on sensitivity to photon noise as a more relevant factor for PL performance. An initial investigation of measurement errors due to photon noise for the Habitable Worlds Observatory pathfinder mission WaveDriver found that a 4-port PL would substantially outperform the Zernike wavefront sensor for low-order aberrations (Gerard \textit{et al.} 2026 \cite{Gerard26Wavedriver}; henceforth G+26). While some configurations of the Zernike are known to be near-optimally sensitive for mid-order aberrations\cite{Chambouleyron21}, this optimality comes at the cost of a large dot diameter, which diminishes sensitivity to lower orders at which the lantern should perform well.

Dynamic range and sensitivity are both relevant to AO/instrument design trade studies; for instance, having a wider dynamic range for the second-stage WFS relaxes constraints on the first stage's performance. However, photon noise sensitivity more directly impacts performance, as it determines its own term in the AO error budget and directly limits the magnitude of the guide star required. 

\subsection{Expectations for PL sensitivity}

A question of interest for PL wavefront sensing is the coupling between sensitivity and mode count. Conventionally, $N$-port PLs are used when they are required to sense at most $N-1$ modes, e.g. Kim \textit{et al.} 2024\cite{Kim24} used a 3-port lantern for tip/tilt sensing, and G+26 used a 4-port lantern for tip/tilt/focus sensing. This is because a lantern with fewer ports will have proportionally more power routed to each port, which should yield maximal sensitivity over a reduced number of aberration modes. However, it is possible that higher port counts could yield similar performance for some modes; for example, two diametrically opposed ports sensing a radially-symmetric focus aberration should be as sensitive as one port at the same radial position. This opens up a trade space for mode count, where sensitivity to higher-order aberrations may be possible without much cost to low-order performance.

Analogous to the previously-mentioned tradeoff between Zernike WFS dot diameter and sensitivity to various spatial orders, we expect PL WFS sensitivity as a function of aberration spatial frequency to trade off against the focal ratio of the input. Smaller f-numbers result in smaller position shifts for the low orders (and hence lower sensitivity to low orders), but smaller f-numbers also yield higher-order patterns that should be seen well by the lantern (hence improving sensitivity to higher modes); the reverse is true for larger f-numbers. Therefore, although sensitivity is set in part by the WFS throughput, the best performance may not necessarily be achieved at the peak throughput. Assessing which configuration yields the best performance may only be possible on a case-by-case basis.

Unlike other WFSs like the pyramid or Zernike, where all or almost all of the light at the input may be used for sensing, the PL should be expected to lose some sensitivity due to its inability to couple in all of the light at its entrance. The large design space of lanterns suggests that such sensitivity losses can be partially mitigated; full control over the linear transformation being applied to the input electric field (likely involving a combination of lantern design optimization and pre- and post-lantern optics) would allow for optimal wavefront sensing. Indeed, analogous work on photonic integrated circuits\cite{Lin26} finds that the sensitivity of an $N$-channel circuit can reach the theoretical limit over a basis of $N-1$ phase aberration modes. It is therefore possible that optimally sensitive lanterns can also be designed, so that the only sensitivity loss comes from the lantern throughput factor. 

\subsection{Computing photon noise sensitivity}

An analysis of WFS sensitivity was first carried out by Guyon \textit{et al.} 2005\cite{Guyon05}. We use the formulation of sensitivity described by Chambouleyron \textit{et al.} 2023\cite{Chambouleyron23}. The sensitivity to photon noise is calculated by measuring the slope of the WFS response to the aberration of interest (as is done when computing an interaction matrix for linear wavefront reconstruction), dividing by the square root of the intensity for a flat wavefront, and taking its norm. Using the notation $I(\phi)$ to mean the WFS intensity in response to the phase screen $\phi$, we have 

\begin{align}
    \label{eq:sensitivity}
    s_\phi = \norm{\frac{\delta I_\phi}{\sqrt{I(0)}}}_2 = \norm{\frac{\frac{I(\epsilon\phi) - I(-\epsilon\phi)}{2\epsilon}}{\sqrt{I(0)}}}_2 = \sqrt{\sum_i \left(\frac{1}{2\epsilon} \frac{I(\epsilon\phi)_i - I(-\epsilon\phi)_i}{\sqrt{I(0)_i}}\right)^2}
\end{align}

Here, $i$ indexes the pixels in the WFS measurement, or in the case of the PL, the summed intensity from each port. This also assumes that the RMS of $\phi$ is 1.

$s_\phi$ is a number between 0 and 2, which is the square root of the Fisher information about the aberration of interest carried by the WFS measurement \cite{Paterson08}. Photon noise sensitivity relates to measurement error in the following way: a measurement with $n_\text{ph}$ photons that is used for linear wavefront reconstruction of an aberration $\phi$ has a variance

\begin{align}
    \label{eq:sensitivity_variance}
    \sigma_\phi^2 = \frac{1}{s_\phi^2 \times n_\text{ph}}.
\end{align}

The Chambouleyron \textit{et al.} 2023 framework is intended for Fourier-filtering WFSs like the pyramid and Zernike WFSs\cite{Fauvarque16}. The PL is not Fourier-filtering, but the same method is applicable as long as we include the throughput loss due to coupling by normalizing to intensity at the pupil plane rather than at the WFS plane. We can quantify this within the sensitivity metric using the relationship to measurement noise. If we have a throughput factor $T$, such that when there are $n_\text{ph}$ photons in the pupil plane we have $T n_\text{ph}$ photons in the WFS plane, the measurement variance is

\begin{align}
    \sigma_{\phi,T}^2 = \frac{1}{s_\phi^2 \times T \times n_\text{ph}}.
\end{align}

Therefore, we can include this in the usual sensitivity measurement if we define $s_{\phi,T} = s_\phi \times \sqrt{T}$. This is a moderately significant factor; for instance, for the PL used in Figure~\ref{fig:optimal_injection}, the maximum throughput being 0.747 implies the peak sensitivity drops from 2 to $2 \times \sqrt{0.747} = 1.729$. While this is a notable drop, a PL sensing at this limit would still outperform, e.g. the unmodulated pyramid WFS, whose sensitivity is $\sqrt{2} \approx 1.414$.

This gives us a quantitative interpretation of our previous expectation of how sensitivity trades off with the focal ratio at the input. Sensitivity will drop with a square root factor of the throughput, which is typically more significant for smaller focal ratios than larger ones, but disincentivizes very large focal ratios as well.

Sensitivity calculations assume perfect linear reconstruction, which is not always justified for the PL. We assume in this case that each aberration is the only one being reconstructed, or equivalently that there is zero crosstalk; this holds in the limit of small aberration amplitudes. For larger aberration amplitudes, we could practically use nonlinear reconstructors\cite{Sengupta24,Lin24fewmode}; this analysis extends to those at least to first order, since their behavior is locally linear. Data-driven methods, like modeling the intensity-to-phase map using a neural network or radial basis functions, effectively interpolate between measured points, meaning they match the linear reconstructor close to a flat wavefront; empirically calibrated methods, like gradient descent over the lantern's transfer matrix, also reduce to linear reconstruction because at a flat wavefront the gradient of the transfer matrix is the interaction matrix\cite{Lin22}. Testing the photon noise sensitivity of a nonlinear reconstructor empirically would be worthwhile but would also be subject to errors from the reconstructors themselves; an easier approach would be to linearize around various points other than a flat wavefront and assume the perfect nonlinear reconstructor would smoothly interpolate between each of these linear regimes. Appendix~\ref{sec:appendix} describes the comparison of ideal photon noise sensitivity to measurement errors due to photon noise from a linear wavefront reconstructor; this comparison should also extend to nonlinear reconstructors. 

\section{Simulations}

\subsection{Simulation methods and common setup}

We simulate PLs using the \textit{lightbeam} Python package\cite{lightbeam} and we simulate the surrounding optics using the \textit{hcipy} Python package\cite{hcipy} as previously detailed in Sengupta \textit{et al.} 2024\cite{Sengupta24}. We combine these in a package called \textit{espeon}\footnote{\href{https://github.com/ucolabforadaptiveoptics/espeon
}{https://github.com/ucolabforadaptiveoptics/espeon
}}, which sets up \textit{lightbeam} runs given lantern design parameters, saves the resulting principal mode basis and pixel-to-grid mapping as HDF5 files, and generates the appropriately-sized grids to simulate the lantern within \textit{hcipy}.

We present simulations for two lantern designs: the 6-port lantern from Kim \textit{et al.} 2022\cite{Kim22}, whose throughput as a function of f-number was shown in Figure~\ref{fig:optimal_injection}; and the 19-port lantern used in the experimental portion of this work from CREOL at the University of Central Florida.\cite{Moraitis2021} In addition, Sengupta \textit{et al.} in these proceedings\cite{Sengupta26oah} shows simulated and empirically-estimated sensitivity curves for two more lantern designs, made at Lawrence Livermore National Laboratory. We consider these two lanterns in this work to see the effect of varying port count within a standard layout. Both lanterns are single-moded at and simulated at 1550nm.

Throughout this work, we consider the PL's sensitivity to sine-wave aberrations where we specify a spatial frequency in cycles per pupil. In simulation, we take a sine wave and a cosine wave and average the two in quadrature, i.e. $s = \sqrt{s_{\sin}^2 + s_{\cos}^2} / \sqrt{2}$. Each individual curve on a photon noise sensitivity plot is the sensitivity as a function of spatial frequency; we plot a range of f-numbers before and after the one showing the optimal throughput.

\subsection{Sensitivity in WFS-only mode}

\begin{figure}[h!]
    \begin{subfigure}{0.49\textwidth}
        \includegraphics[width=\linewidth]{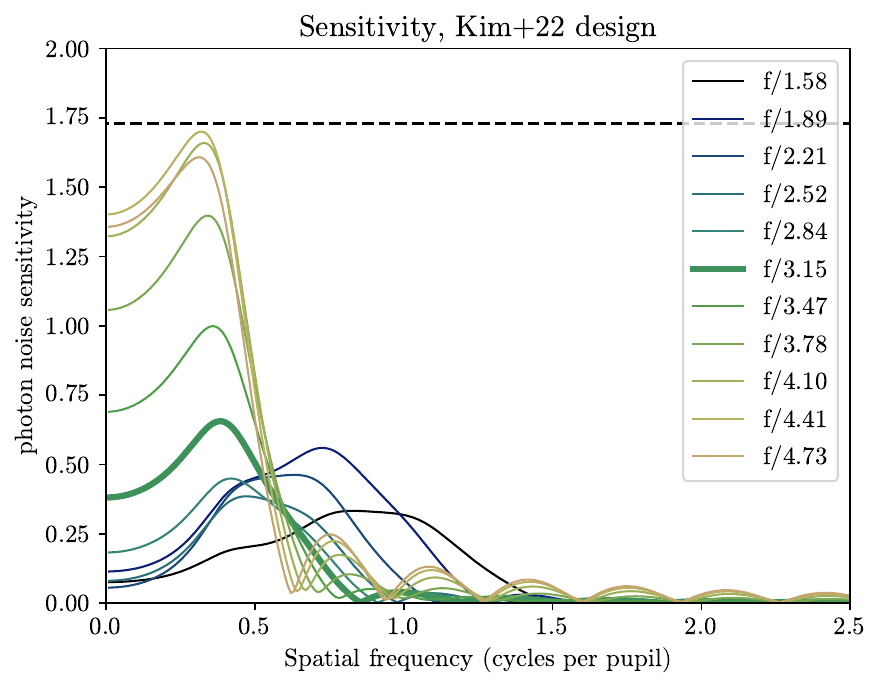}
        \caption{K+22 6-port PL.}
        \label{fig:sensitivity_kim22}
    \end{subfigure}\hspace*{\fill}
    \begin{subfigure}{0.49\textwidth}
        \includegraphics[width=\linewidth]{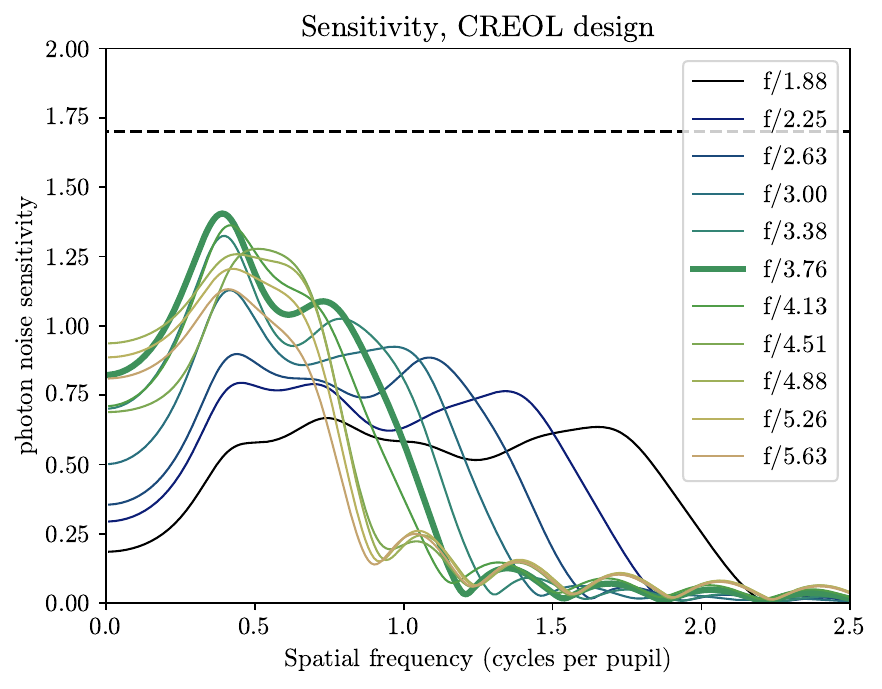}
        \caption{CREOL 19-port PL.}
        \label{fig:sensitivity_creol}
    \end{subfigure}
    \caption{Sensitivity as a function of spatial frequency for both lantern designs. The thicker green line represents the f-number with the optimal throughput, and the dotted black line shows the throughput-adjusted fundamental limit, equal to 2 times the square root of the optimal throughput.}
    \label{fig:sensitivity_sim_wfsonly}
\end{figure}

Figure~\ref{fig:sensitivity_sim_wfsonly} shows the sensitivity to photon noise for the two lantern designs. 

The K+22 6-port PL is sensitive only to lower-order aberrations, with very low sensitivity past 1 cycle per pupil; this is expected as it does not have enough ports to sense higher-order patterns. The K+22 PL is also significantly closer to its fundamental limit for a low-order mode around 0.4 cycles per pupil when overfilling the input by around 1.4x relative to the peak; this configuration would likely mostly couple in the PSF core, so we observe enhanced sensitivity to shifts in that core at the cost of higher-order patterns. The peak throughput f-number for this lantern largely underperforms some of the higher f-numbers. 

The CREOL 19-port PL is sensitive to relatively higher-order aberrations due to its higher port count. We see a clear tradeoff between the spatial frequency and peak sensitivity, as previously anticipated; underfilling the input allows for more higher-order patterns to couple into the lantern, and within the range considered there are sufficient ports to capture these signals. However, this is at the cost of diminished throughput, causing the peak sensitivity to be lower. In this case, the highest sensitivity overall is at the f-number with the optimal throughput, and overfilling does not show much benefit. The highest sensitivity is significantly below the peak based on throughput; this may be a result of the light being split up across a larger number of ports.

We note the higher-order ringing seen in both lanterns; notably, this does not appear in throughput curves, suggesting this is a result of relative periodicity between the aberration and the size of the multi-mode end.

\subsection{Sensitivity in dual imager/WFS mode}

To simulate dual use as an imager and wavefront sensor, we recompute sensitivities with two ports removed, assuming that they are being used for science purposes only, such as imaging. This choice is somewhat simplified and could be made more specific based on the observing scenario. One limitation of this choice is that it is not necessarily the case that improving the wavefront on the remaining $N-2$ ports will improve the relevant scientific figure of merit. In the worst case, where all aberration modes couple into their own ports or into only ports in the subset designated for one purpose or the other, we would strictly improve the wavefront on the WFS subset and would not see any improvement for science imaging. These results are still likely to be relevant since this worst case is unlikely for low-order aberrations in a standard lantern, and this concern could also be mitigated if we were to use the science ports for image correction in postprocessing. We choose two ports assuming our science case is nulling interferometry\cite{Xin24,Diaz24,Diaz26}; we always remove the central port as it carries on-axis starlight, and we remove one other port corresponding to the planet location. Due to radial symmetry in the simulations, all of the remaining possible cases reduce to three: for the K+22 design, removing any other port (there is only one ring of ports around the central one); for the CREOL design, removing any port from the inner ring; and for the CREOL design, removing any port from the outer ring.

\begin{figure}[h!]
    \begin{subfigure}{0.32\textwidth}
        \includegraphics[width=\linewidth]{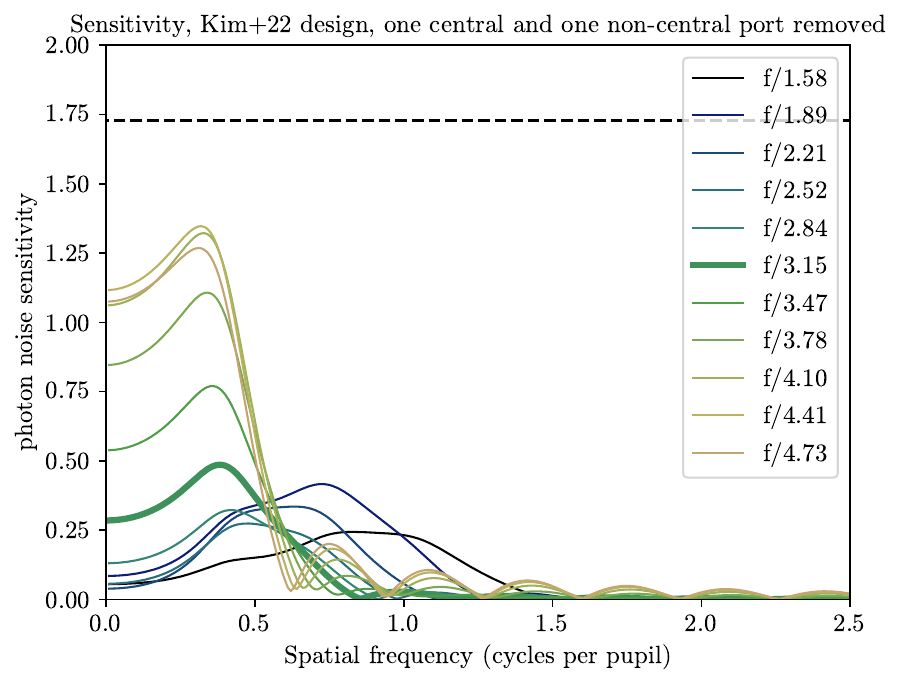}
        \caption{K+22 6-port PL, one central and one non-central port removed.}
        \label{fig:sensitivity_kim22_twoout}
    \end{subfigure}\hspace*{\fill}
    \begin{subfigure}{0.32\textwidth}
        \includegraphics[width=\linewidth]{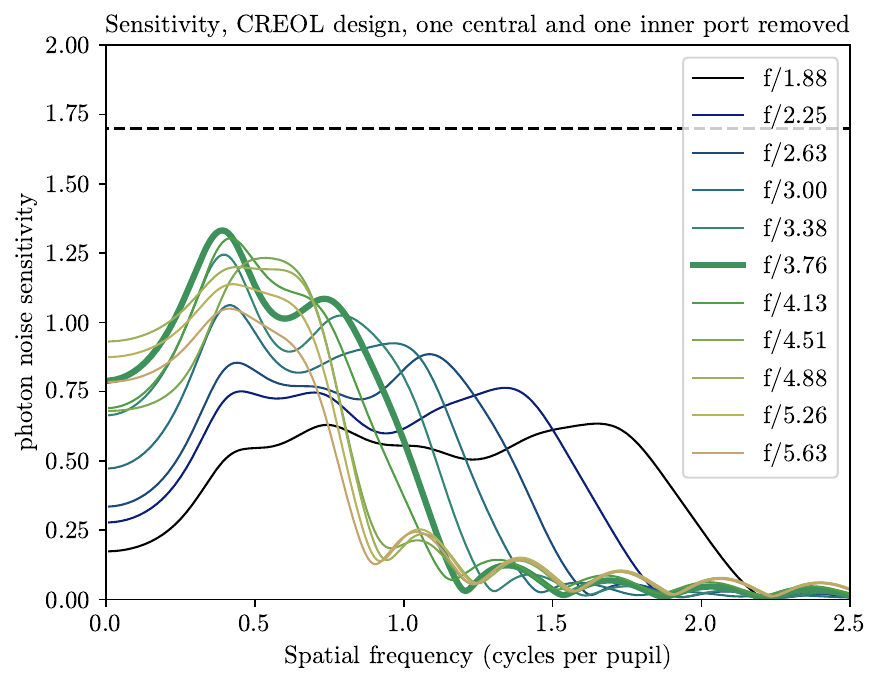}
        \caption{CREOL 19-port PL, one central and one inner port removed.}
        \label{fig:sensitivity_creol_twoout_inner}
    \end{subfigure}
    \begin{subfigure}{0.32\textwidth}
        \includegraphics[width=\linewidth]{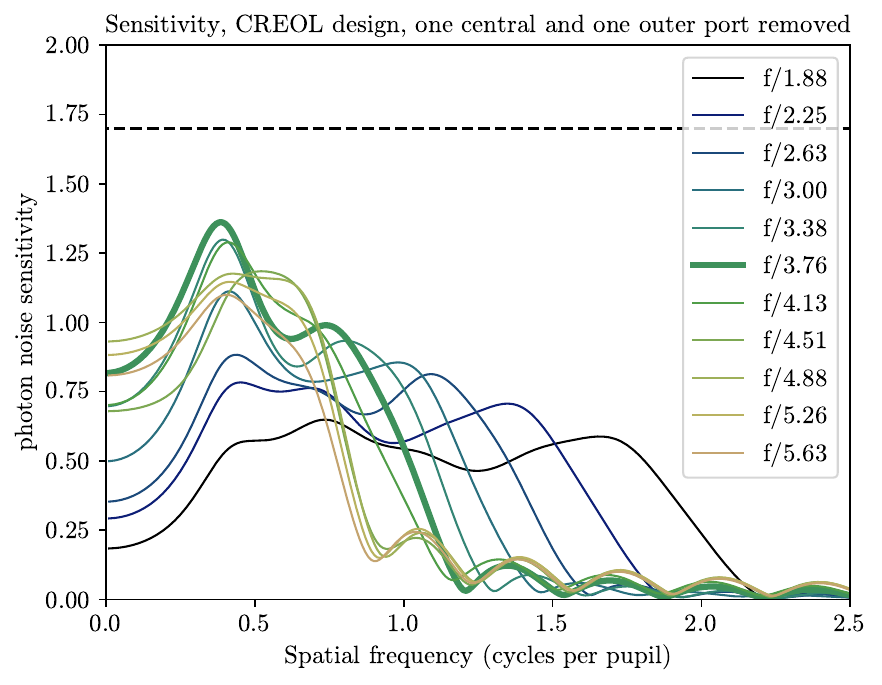}
        \caption{CREOL 19-port PL, one central and one outer port removed.}
        \label{fig:sensitivity_creol_twoout_outer}
    \end{subfigure}
    \caption{As in Figure~\ref{fig:sensitivity_sim_wfsonly}, but with two ports removed in each case.}
    \label{fig:sensitivity_sim_wfsonly_removed}
\end{figure}

We note that while sensitivity reduces in all cases, the reduction is more significant for the 6-port PL than the 19-port PL, reflecting that a larger proportion of the light being used for WFS was removed. The highest sensitivity for the K+22 lantern is no longer close to the theoretical maximum, whereas the highest sensitivity for the two CREOL cases is almost unchanged. Cases in which the removed port is in the inner ring show slightly better sensitivity to higher-order aberrations (around 2 cycles per pupil) than when the removed port is in the outer ring. This reflects where in the lantern the higher-order patterns couple in.

\section{Experimental setup and results}

\subsection{Laboratory procedure}
We use the muirSEAL testbed\cite{Sengupta25} to measure the photon noise sensitivity of a 19-port PL from CREOL (the same lantern as in the initial muirSEAL work).

\begin{figure}
    \includegraphics[width=\textwidth]{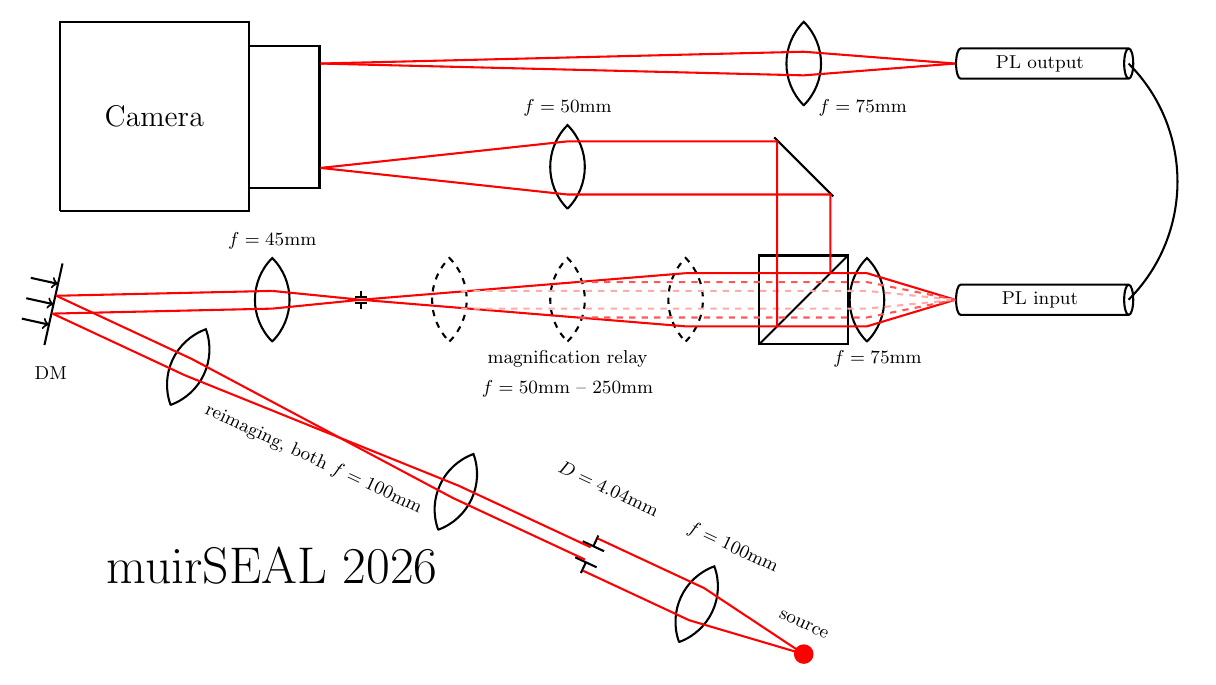}
    \caption{Updated muirSEAL configuration. Compared to the previous design, the PL input stays fixed, and the f-number is changed by varying the beam diameter via changing the magnification.}
    \label{fig:muirseal_diagram}
\end{figure}

muirSEAL's initial design was intended to test PL wavefront sensing performance at a range of input focal ratios. We make one significant design change, shown in the updated configuration in Figure~\ref{fig:muirseal_diagram}. In the initial design, the lens immediately before the lantern was switched out, and the lantern was brought to a focus with respect to that lens using a travel stage. However, in that design, small angle errors on the beam after the DM that were too small to correct during alignment were amplified over the long focal length, causing significant position offsets and suboptimal coupling into the lantern. Further, a calculation error meant this design actually yielded f-numbers that were significantly larger than expected. We fix this by placing a lens with the smallest available focal length after the DM, and changing the lens that re-collimates the beam in order to change the magnification. Table~\ref{tab:muirseal_params} lists the focal lengths used and the corresponding focal ratios. The rounded focal ratios are used as labels in plots.

\begin{table}[h!]
    \centering
    \begin{tabular}{|p{3.75cm}|p{3.75cm}|p{3.75cm}|p{3.75cm}|}
        \hline
        \textbf{Magnifying lens focal length (mm)} & \textbf{Diameter at beam cube (mm)} & \textbf{Focal ratio} & \textbf{Focal ratio (rounded)} \\\hline
        50 & 4.49 & 16.71 & 17\\
        75 & 6.73 & 11.14 & 11\\
        100 & 8.98 & 8.35 & 8 \\
        150 & 13.47 & 5.57 & 6\\
        200 & 17.96 & 4.18 & 4\\
        250 & 22.44 & 3.34 & 3\\
        \hline
    \end{tabular}
    \caption{Magnifying lens choices and corresponding focal ratios.}
    \label{tab:muirseal_params}
\end{table}

The magnifying lens is mounted within an optical cage, keeping its x/y position constant. We manually replace and move this lens to create each f-number at the lantern input; minor adjustments to the lantern's tip/tilt are required, but not to the lantern's focus position. We were not able to carry out detailed measurements of throughput as a function of f-number, but we observed based on the flat-wavefront measurements at each f-number that the best throughput was at f/8. This is not consistent with the design and may be an experimental artifact; further analysis of the optimal coupling at each f-number is needed to confirm this. A more full comparison between f-numbers and across WFSs would require a stable alignment for each f-number individually and increased optimization of coupling into the PL. Such alignment has proven difficult so far.

We induce aberrations using the deformable mirror (DM). We generate sine-wave aberrations with spatial frequencies in the range 0.1-2.5 cycles per pupil; since we only illuminate a 5-segment wide region of the DM out of a total of 15 segments across, this is equivalent to 3 times the spatial frequency across the full DM. We make these phase screens in \textit{hcipy} and project them onto a basis of influence functions formed using the \textit{SegmentedDeformableMirror} class to generate commands for the DM. We take measurements at 4 sine-wave rotation angles (0, 30, 60, 90 degrees), and at amplitudes in the range -1 to +1 rad to check for linear ranges. We repeat this for the six focal ratios. The relevant code for operating the testbed is in a package called \textit{umbreon}\footnote{\href{https://github.com/ucolabforadaptiveoptics/umbreon}{https://github.com/ucolabforadaptiveoptics/umbreon}}.

As with previous muirSEAL experiments, we sum the counts in an aperture containing each port in turn to generate the WFS readouts as 19-length arrays. We choose one poke amplitude per f-number to generate interaction matrices, and use them to compute photon noise sensitivity as in the simulations. We normalize sensitivity measurements (i.e. generate the $\sqrt{I(0)}$ term in Equation~\ref{eq:sensitivity}) based on the readout for a flat wavefront at that f-number for an equivalent comparison, but we further normalize the resulting sensitivity measurements based on the flat-wavefront measurement at f/8, i.e. we multiply all sensitivity measurements by a factor of (total counts for flat wavefront at this f-number)/(total counts for flat wavefront at f/8). This partially accounts for the required throughput factor in sensitivity, but we were not able to fully normalize this to the total power at the pupil plane (PSF measurements were not available for these data due to alignment difficulties.) These measurements should therefore be considered upper limits and should not be directly compared with other WFSs.

\subsection{Measurements in WFS-only mode}

\begin{figure}[h!]
    \includegraphics[width=\textwidth]{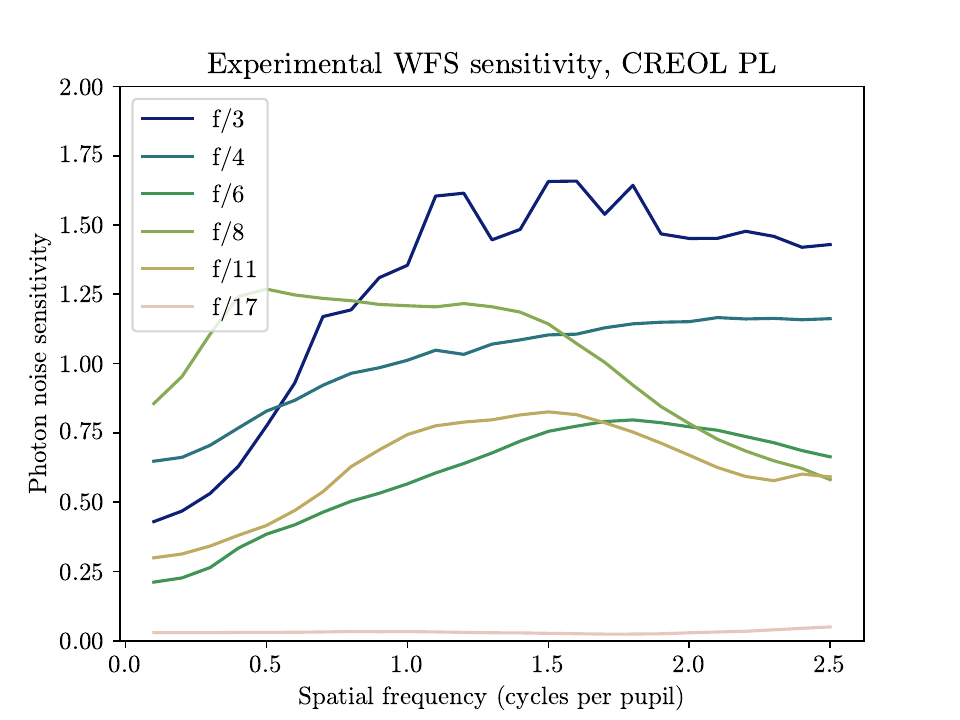}
    \caption{Photon noise sensitivity measurements from muirSEAL (not normalized to power at the PL entrance.)}
    \label{fig:muirseal_sensitivity}
\end{figure}

Figure~\ref{fig:muirseal_sensitivity} shows the experimentally measured photon noise sensitivity curves. f/3 is the most sensitive overall, and peaks at a relatively high spatial frequency. f/4 and f/6 also peak at high spatial frequencies, with f/8 and f/11 appearing to peak at lower spatial frequencies. Although the trend is not monotonic, with f/11 peaking at around 1.5 cycles per pupil and f/8 peaking at around 0.5 cycles per pupil, this largely follows what we expect: smaller focal ratios are broadly sensitive to higher spatial frequencies, as higher-order PSF patterns can couple into the lantern better. We do not observe a strong trend in the magnitude of the peak amplitude with f-number, which is likely due to the differences in overall throughput from having to realign for each f-number. Here, since we normalize to the highest throughput in the PL plane, the appropriate fundamental limit to compare to is 2; scaled relative to this, f/3 is comparable to or better than in simulation. f/17 shows almost no sensitivity due to having much worse coupling efficiency than the other cases. 

\subsection{Measurements in dual imager/WFS mode}

\begin{figure}[t!]
    \begin{subfigure}{0.49\textwidth}
    \includegraphics[width=\linewidth]{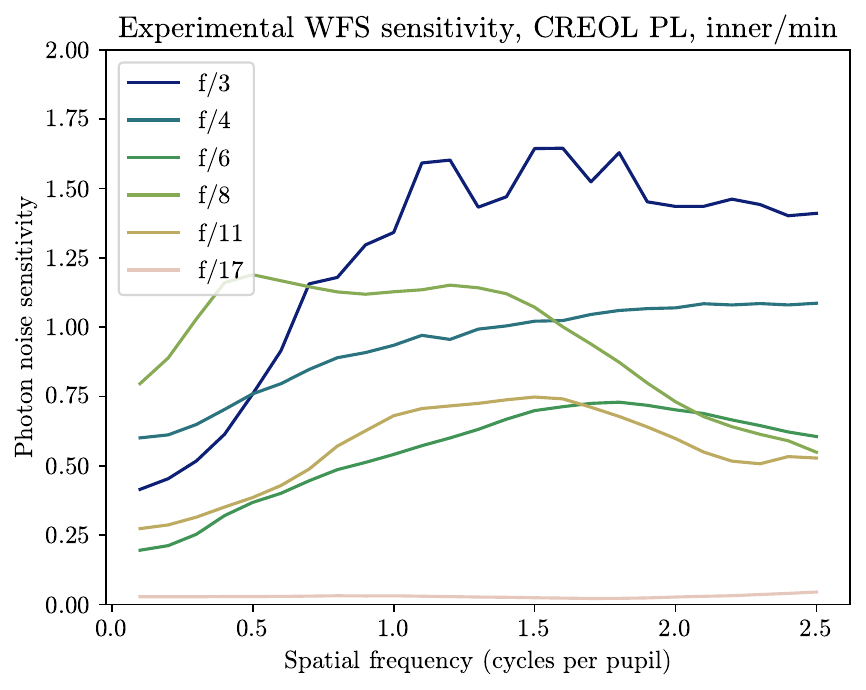}
    \caption{Inner ring, minimum sensitivities} \label{fig:muirseal_sensitivity_lto_inner_min}
    \end{subfigure}\hspace*{\fill}
    \begin{subfigure}{0.49\textwidth}
    \includegraphics[width=\linewidth]{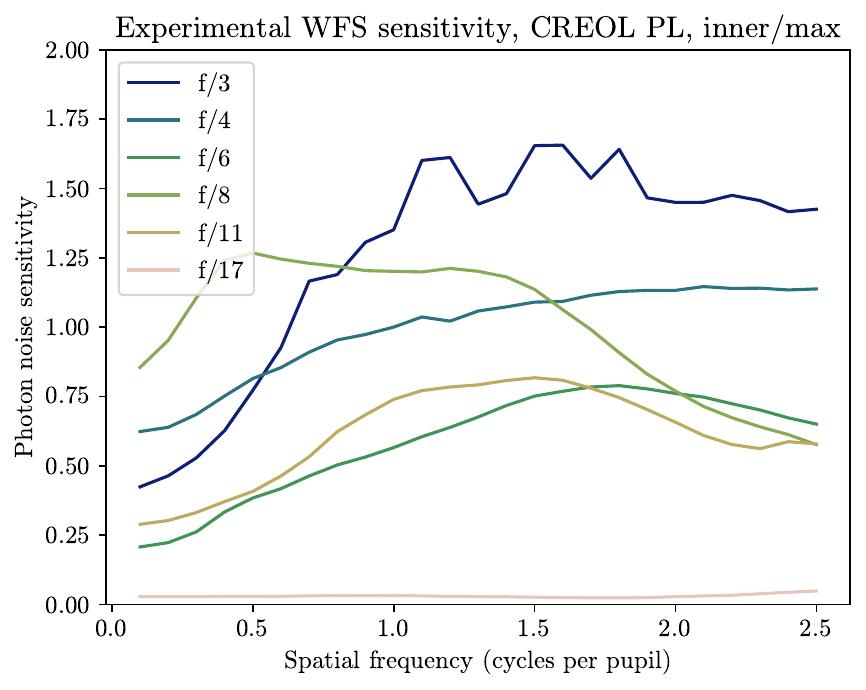}
    \caption{Inner ring, maximum sensitivities} \label{fig:muirseal_sensitivity_lto_inner_max}
    \end{subfigure}\hspace*{\fill}

    \medskip
    \begin{subfigure}{0.49\textwidth}
    \includegraphics[width=\linewidth]{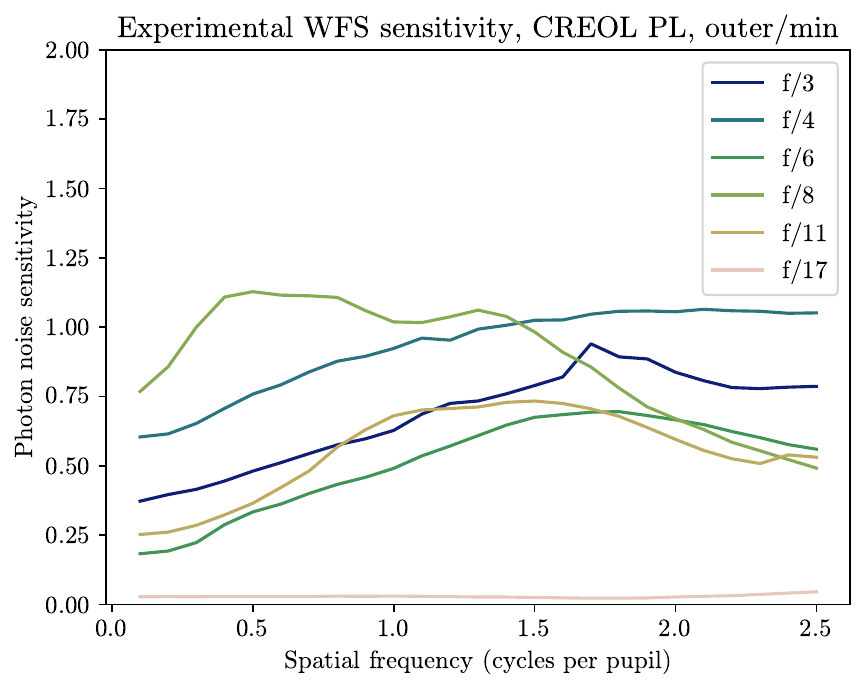}
    \caption{Outer ring, minimum sensitivities} \label{fig:muirseal_sensitivity_lto_outer_min}
    \end{subfigure}\hspace*{\fill}
    \begin{subfigure}{0.49\textwidth}
    \includegraphics[width=\linewidth]{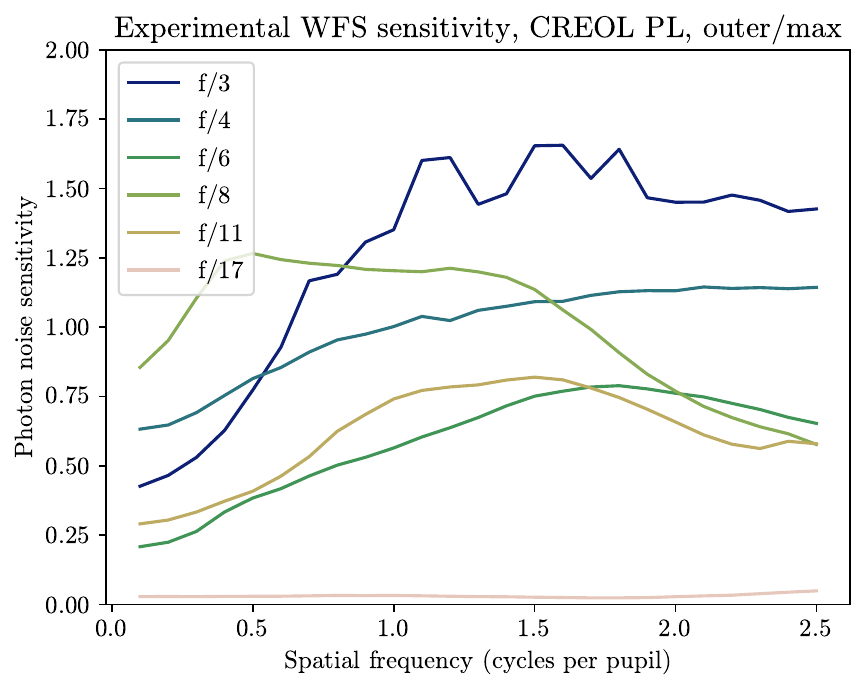}
    \caption{Outer ring, maximum sensitivities} \label{fig:muirseal_sensitivity_lto_outer_max}
    \end{subfigure}\hspace*{\fill}
    
    \caption{``Leave-two-out'' sensitivity measurements. At each point, the port whose exclusion would minimize (for (a) and (c)) or maximize (for (b) and (d)) the sensitivity is left out. In (a) and (b) the chosen port is restricted to the inner ring (ports 2-7) and in (c) and (d) it is restricted to the outer ring (ports 8-19).}
    \label{fig:muirseal_sensitivity_lto}
\end{figure}

Since on a real lantern we cannot assume radial symmetry as we did in simulation, we need to consider the impact of leaving out all 18 of the non-central ports in addition to the central port. We summarize these 18 cases by plotting the worst and best sensitivity pointwise to see the range of possible outcomes. We separate out cases where the excluded port is from the inner ring vs. the outer ring.

Figure~\ref{fig:muirseal_sensitivity_lto} shows the resulting extremal sensitivity curves. We see very few changes other than the minimum f/3 curve when removing an outer port; this suggests that other than that individual outlier, our measurements are not overly reliant on any one port, and WFS performance should be largely similar when any two ports are removed for science imaging.

\section{Conclusion}


We have carried out the first simulations and experimental measurements of photon noise sensitivity for the photonic lantern wavefront sensor. This is a metric that will substantially determine performance for systems incorporating the PL as a second-stage WFS. We find in simulation that the PL's performance for low-order aberrations is favorable, sometimes reaching the limit set based on its total throughput. We find experimentally that the PL shows roughly the expected trends as a function of focal ratio at the input, and that it shows relatively high sensitivity to photon noise overall.

Future laboratory experiments will optimize coupling at each f-number to a greater degree and will more accurately baseline sensitivity measurements based on observed throughput. Further tests could also include a wider range of configurations, e.g. lanterns with fewer ports, multiple lanterns with the same design, or applying fixed beam-shaping offsets with the DM. It would also be relevant to compare `true' interaction-matrix sensitivities, `empirical' sensitivities based on the variance of wavefront reconstruction results, and `modeled' sensitivities based on expected performance from lantern characterizations, e.g. using off-axis holography.

This work sets the stage for more integrative instrument trade studies involving the PL as a joint science instrument and WFS. Just within the WFS component, there are many elements that can be traded off, only a few of which were considered here. These include pre-lantern optics (beam-shaping/apodization, input focal ratio), the lantern design (number of ports, positioning, spacing), and post-lantern optics (filtering, dispersion, injection into downstream devices). We have shown that by controlling only the input focal ratio and the number of ports, we can achieve a wide range of performance outcomes; to illustrate an additional dimension, G+26 found that for the same number of photons, idealized monochromatic performance and dispersed performance over a $\sim$90nm bandpass were almost the same.

Future instrument designs will have to choose all of these factors not only to maximize sensitivity to photon noise, but to maximize overall scientific return; sensitivity must be balanced with dynamic range and with the number of modes that need to be corrected. For instance, a design could choose to slightly under- or over-inject relative to the peak throughput, because the additional sensitivity on particular modes is enough to compensate for the throughput loss when in closed loop, and the modes that are not sensed as well as a result are known not to have large  residuals from the first stage. This is additionally complicated by the need for specific scientific figures of merit. 

Overall, we find that the PL as a low-order WFS is competitive with well-established choices in terms of photon efficiency.

\acknowledgments     
A.S. thanks Rachel Morgan, Jonathan Lin, Parth Nobel, and Jules Fowler for valuable discussions. This work was funded in part by the Heising-Simons Foundation (grant \#2020-1822). 

\appendix
\section{Checking the sensitivity metric against simulated measurement errors}
\label{sec:appendix}
To verify our implementation of the sensitivity metric, we check it against simulations of linear wavefront reconstruction with Poisson noise added (`simulated empirical sensitivity'). This metric is similar to that used in G+26, except in that case the measurement error with respect to the number of photons was translated to a guide star magnitude rather than to sensitivity. We carry out this simulation for the unmodulated pyramid WFS, since its true sensitivity for almost all modes should be $\sqrt{2}$\cite{Chambouleyron23}.

We compute the sensitivity metric as described earlier in this work. We simulate WFS images using the \texttt{PyramidWavefrontSensorOptics} class in \textit{hcipy} (taking the intensity of the image in the WFS plane and multiplying by the grid weights), and we normalize images by dividing by the total power in the pupil-plane wavefront.


\begin{figure}[b!]
    \centering
    \includegraphics{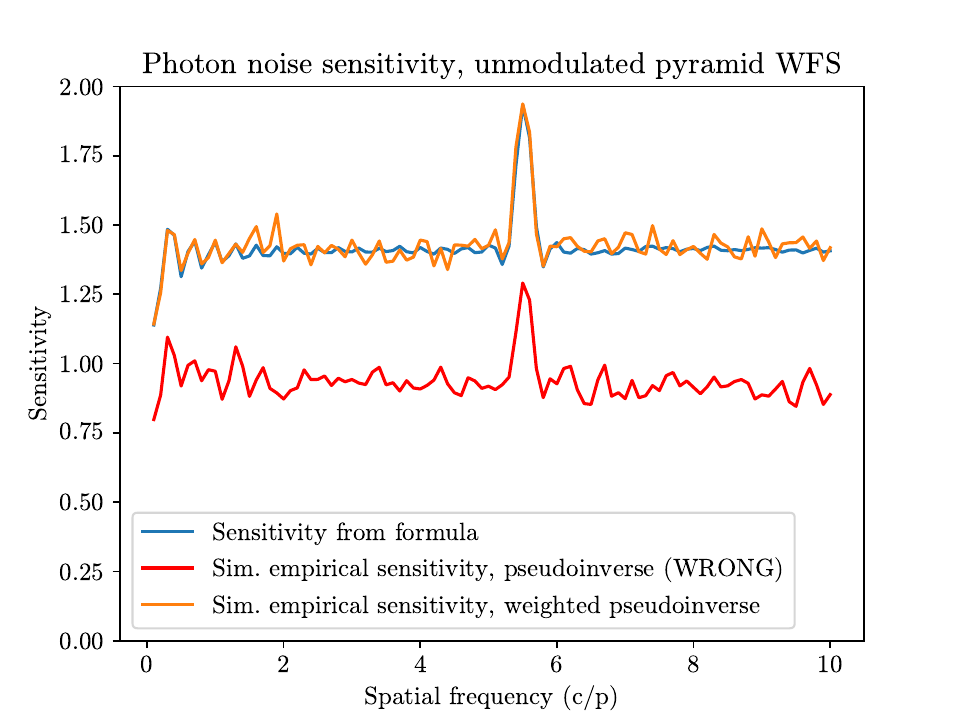}
    \caption{The photon noise sensitivity of the unmodulated pyramid WFS computed three ways (one incorrectly to illustrate the difference): (1) from the standard formula; (2) empirically from averaging reconstruction results in the presence of photon noise, but using the incorrect estimator; and (3) as with (2) but using the correct estimator. We see that (1) and (3) line up closely and that (2) underestimates the true sensitivity significantly.}
    \label{fig:weighted_pinv_pywfs}
\end{figure}

To calculate simulated empirical sensitivity, we are required to take a weighted pseudoinverse of the interaction matrix. The reason for this is we assume only photon noise is present, meaning the normal pseudoinverse (calculated using, e.g. \texttt{numpy.linalg.pinv} or \texttt{hcipy.util.inverse\_tikhonov} on the interaction matrix) does not give us the maximum-likelihood estimator of the phase; this would only hold if the noise was Gaussian. A statistical argument for this is provided in the appendix of Lin 2026\cite{Lin26}; we slightly expand on the most relevant portion for this analysis. Suppose we have an aberration $\phi$ and we obtain a measurement $I(\phi) + n$, where $n$ represents a noise term. The Cramer-Rao lower bound for $\hat{\phi}$, i.e. the best possible reconstruction of $\phi$ given such a measurement, is given by

\begin{align}
    \text{Cov}(\hat{\phi}) \succeq \mathcal{I}(\phi)^{-1}= (M^\intercal \Sigma^{-1} M)^{-1}
\end{align}

where $\mathcal{I}$ represents Fisher information and $M$ is the interaction matrix. The diagonal of the Fisher information matrix is related to the sensitivity metric.

The reconstructor that reaches the Cramer-Rao lower bound is the maximum-likelihood estimator of $\phi$. For a measurement $y$, this estimator is $\hat{\phi} = M^\dagger y$ where 

\begin{align}
    M^\dagger = (M^\intercal \Sigma^{-1} M)^{-1} M^\intercal \Sigma^{-1}.
\end{align}

If the noise were Gaussian, we would have $\Sigma$ equal to the identity, and this would reduce to the usual formula for the Moore-Penrose pseudoinverse. However, in this case, $\Sigma$ is the WFS-plane intensity for a flat wavefront (since the mean and variance of a Poisson distribution are the same), so we instead take a weighted pseudoinverse. For each simulated noiseless WFS image, we add photon noise, normalize the result, and reconstruct the wavefront using the noisy image. We take 500 realizations of this process and compute the empirical sensitivity from the variance of these realizations using Equation~\ref{eq:sensitivity_variance}.

Figure~\ref{fig:weighted_pinv_pywfs} demonstrates the difference between applying and not applying this weighting. The empirical calculation lines up with the standard one when the weighting is used, and is significantly off otherwise.

\bibliography{report} 
\bibliographystyle{spiebib} 

\end{document}